# Elastic isotropy of ε-Fe under Earth's core conditions

Xianwei Sha, and R. E. Cohen

Carnegie Institution of Washington, 5251 Broad Branch Road, NW, Washington, D. C. 20015, USA

**Abstract**

Our first-principles calculations show that both the compressional and shear waves of ε-Fe become elastically isotropic under the high temperatures of Earth's inner core conditions, with the variation in sound velocities along different angles from the c axis within 1%. We computed the thermoelasticity at high pressures and temperatures from quasiharmonic linear response linear-muffin-tin-orbital calculations in the generalized-gradient approximation. The calculated anisotropic shape and magnitude in ε-Fe at ambient temperature agree well with previous first-principles predictions, and the anisotropic effects show strong temperature dependences. This implies that other mechanisms, rather than the preferential alignment of the ε-Fe crystal along the Earth's rotation axis, account for the seismic P-wave travel time anomalies. Either the inner core is not hcp iron, and/or the seismologically observed anisotropy is caused by inhomogeneity, i.e. multiple phases.

**I. Introduction**

The Earth's inner core, mainly consisting of solid iron, is elastically anisotropic, where the seismic waves travel ~3-4% faster parallel to the Earth's spin axis than parallel to the equatorial plane [*Morelli, et al.*, 1986; *Romanowicz, et al.*, 1996; *Song*, 1997; *Tromp*, 1993; *Woodhouse, et al.*, 1986]. This seismic anisotropy is believed to be associated with the preferential alignment of the iron crystals composing the inner core [*Hemley and Mao*, 2001; *Karato*, 1999; *Song*, 1997; *Stixrude and Cohen*, 1995a; b]. The hexagonal-close-packed phase of iron (ε-Fe) is reported to have a wide stability field including at the extremely high temperature (4000 to 8000 K) and pressure (330 to 360 GPa) found in the Earth's deep inner core [*Hemley and Mao*, 2001]. Nuclear inelastic x-ray scattering experiments in a laser-heated diamond anvil cell show that the sound velocities in ε-Fe change significantly with increasing temperature, and that Birch's law, which assumes a linear relation between the sound velocities and atomic densities, fails at high pressures and temperatures [*Lin, et al.*, 2005]. Their measurements are limited to 73 GPa and 1700 K, significantly below inner core conditions. Steinle-Neumann et al. used first-principles calculations and the particle-in-cell model to examine the thermoelasticity at the core [*Steinle-Neumann, et al.*, 2001], but their calculations gave too large *c/a* axial ratios at high temperatures [*Sha and Cohen*, 2006b]. Vocadlo computed the sound velocities of iron and several iron alloys via *ab initio* finite temperature molecular dynamics simulations, but ε-Fe results were obtained at only two high temperatures and for limited supercell sizes due to the intensive computational costs [*Vacadlo*, 2007]. Here we reexamine the sound velocity and elastic anisotropy of hcp-Fe at high temperatures and pressures, including under core conditions, based on first-principles calculated thermal equation of state and elastic moduli.

**II. Results and discussion**

For both the longitudinal wave and two shear waves with polarization orthogonal to and inside the vertical plane, the sound velocities of hcp crystals such as ε-Fe depend on the elastic moduli and the angle between the 6-fold axis and the propagation direction [*Royer and Dieulesaint*, 2000]. We obtained the elastic moduli as the second derivatives of the Helmholtz free energies with respect to strain tensor, by applying volume-conserving strains and relaxing the symmetry-allowed internal coordinates. We calculated the Helmholtz free energy as a function of volume, temperature, and strain, including the electronic excitation contributions from the band structures and lattice vibrational contributions from quasi-harmonic lattice dynamics [*Sha and Cohen*, 2009]. The computational approach is based on the density functional theory and density functional perturbation theory, using multi-$\kappa$ basis sets in the full-potential linear-muffin-tin-orbital (LMTO) method [*Savrasov*, 1996; *Savrasov and Savrasov*, 1992]. We used the Perdew-Burke-Ernzerhof (PBE) generalized-gradient-approximation (GGA) for the exchange and correlation energy [*Perdew, et al.*, 1996]. We chose 4-6 values for each strain ranging from 0 to ±0.03, and performed first-principles linear response calculations to obtain the electronic and phonon density of states for all the strained structures at each volume. We then calculated the Helmholtz free energies at temperatures from 0 to 6000 K, and fitted a polynomial of the free energies to get the elastic moduli as the coefficients of the quadratic terms in strain, which appear in the equations of motion and directly give sound velocities [*Barron and Klein*, 1965; *Gregoryanz, et al.*, 2000; *Sha and Cohen*, 2009; *Steinle-Neumann, et al.*, 1999].

We show the calculated elastic anisotropy for ε-Fe at ambient temperature and 50 GPa in Fig. 1. For both the compressional and shear waves that propagate along different angles from c axis, our full potential LMTO sound velocities agree within 1.5% with previous linearized-augmented-plane-wave (LAPW) results [*Steinle-Neumann, et al.*, 1999; *Steinle-Neumann, et al.*, 2001]. At ambient temperature, the first-principles calculations predict a sigmoidal shape for the compressional waves with maximal velocity along the c-axis, and a "bell-like" shape for the shear wave with maximal velocity at $45^\circ$ from the basal plane [*Cohen, et al.*, 1997; *Soderlind, et al.*, 1996; *Steinle-Neumann, et al.*, 1999; *Steinle-Neumann, et al.*, 2001; *Stixrude and Cohen*, 1995b; *Vocadlo, et al.*, 2003]. On the contrary, the radial x-ray diffraction measurements on polycrystalline ε-Fe samples give totally different anisotropic shapes [*Antonangeli, et al.*, 2004b; *Mao, et al.*, 1998; 1999; *Merkel, et al.*, 2005; *Singh, et al.*, 1998]. The discrepancy between theory and experiment is in contrast to another hcp transition metal cobalt, where the experiment [*Antonangeli, et al.*, 2005; *Antonangeli, et al.*, 2004a; *Crowhurst, et al.*, 2006] and theory [*Steinle-Neumann, et al.*, 1999] both give the sigmoidal shape for the compressional wave anisotropy. However, it has now been shown that the assumption of a single uniform macroscopic stress applied to all grains is violated for hcp transition metals under nonhydrostatic compression due to plastic deformation in a sample with preferred orientation, so the single-crystal elastic moduli extracted from polycrystalline radical x-ray diffraction experiments contain large errors [*Antonangeli, et al.*, 2006; *Merkel, et al.*, 2006]. For hcp cobalt, the measured $C_{11}$, $C_{33}$, $C_{12}$ and $C_{13}$ using polycrystalline samples and the same techniques as ε-Fe measurements are 20% off with respect to single-crystal measurements, and the discrepancies are up to 50% and 300% for shear moduli $C_{66}$ and $C_{44}$ [*Antonangeli, et al.*, 2006]. For the longitudinal sound velocity as a function of propagation direction with respect to the c-axis in hcp cobalt at high pressure, although the polycrystalline-sample shape agrees well with ε-Fe experimental data, the elastic anisotropy obtained based on single-crystal moduli has the same shape as first-principles predictions [*Steinle-Neumann, et al.*, 1999], which is similar to our calculated ε-Fe shape. Therefore, we have confidence in our first-principles elasticity computations of anisotropy.

The isotropic aggregate sound velocities in solids are related to the elastic moduli according to the Christoffel equation [*Gulseren and Cohen*, 2002; *Sha and Cohen*, 2006a]. The calculated sound velocities decrease with increasing temperature and decreasing atomic density, consistent with nuclear inelastic x-ray scattering experiment [*Lin, et al.*, 2005]. We calculated the sound velocities along the Hugoniot based on our first-principles shock equation of state and elastic moduli, in comparison to recent shock-compression measurements [*Nguyen and Holmes*, 2004]. Along the Hugoniot curve, the calculated compressional wave velocities agree with the shock data up to 220 GPa, and the bulk velocities match with the experiment after 250 GPa, as shown in Fig. 2. The experiment suggests a solid-liquid phase transition along the Hugoniot at 225±3 GPa, and the liquid sound velocities are known to be very close to the solid bulk velocities [*Gulseren and Cohen*, 2002].

We show the temperature dependences of the anisotropic effects in ε-Fe at 48 bohr$^3$/atom in Fig. 3. With increase in temperature, the elastic anisotropy decreases dramatically. At the Earth's inner core conditions (~6000 K), our calculations predict that both the compressional and shear waves are elastically isotropic, where the sound velocities along different propagation direction agree within 1%. This is in significant contrast to two previous first-principles calculations which both predicted ~6-10% compressional-wave anisotropy with maximal sound velocity along 90 degree from the c axis for ε-Fe under core conditions [*Steinle-Neumann, et al.*, 2001; *Vacadlo*, 2007]. Steinle-Neumann et al.'s calculations gave too large c/a ratios at high temperatures [*Sha and Cohen*, 2006b]. We interpolated our moduli to obtain the elastic properties at the two temperatures and atomistic densities that Vocadlo examined for ε-Fe. In spite of significantly different predicted elastic anisotropy, the isotropic aggregate sound velocities agree within 2%. For the individual moduli, $C_{13}$ and $C_{33}$ agree in ~5%, $C_{11}$ and $C_{12}$ agree within ~10%. The large discrepancies come from the shear moduli $C_{66}$ and $C_{44}$, with differences up to 15% and 35%. It should be noted that the calculated zero-temperature shear moduli agree well [*Vocadlo, et al.*, 2003], so the discrepancies must come from the thermal parts. Different theoretical techniques have been used to obtain the thermal contributions. We use linear response lattice dynamics, and Vocadlo used *ab initio* molecular dynamics and thermodynamic integration. One major difference is that our calculations are based on quasi-harmonic approximations. However, as shown in earlier calculations using both thermodynamic integration and the PIC model [*Alfe, et al.*, 2001; *Sha and Cohen*, 2006b], the on-site anharmonicity in ε-Fe is small up to the melting temperature. Other possible sources for the discrepancies might be the different set-ups in the first-principles calculations. Vocadlo used a 64-atom supercell and 4 irreducible *k* points in her *ab initio* molecular dynamics simulations. We carefully compare the calculated $C_{44}$ values at different *k* point meshes up to 24×24×24 and *q* meshes up to 6×6×6, and make sure our results are well converged. The Raman-active $E_{2g}$ phonon correlates with the zone–edge acoustic mode, the elastic modulus $C_{44}$, and shear-wave velocity, and their frequencies at high pressures have been measured using the Raman spectroscopy [*Goncharov and Struzhkin*, 2003; *Merkel, et al.*, 2000]. Raman measurements at high temperatures and pressures might be helpful to understand the temperature dependences of the shear moduli and resolve the discrepancy between the two theoretical calculations.

There are several possible geophysical explanations to account for the differences between the predicted elastic isotropy of ε-Fe under core conditions and the observed seismic P-wave travel time anomalies. One possibility is that the seismic anisotropy could not be well explained by these first-principles data since the inner core is not under constant density or isothermal conditions. According to the Preliminary Reference Earth Model (PREM), the inner-core density varies from 47.6 bohr$^3$/atom at zero radius to 49 bohr$^3$/atom at the inner-outer core boundary [*Ahrens*, 1995]. The adiabatic temperature difference is estimated to be up to 350 K from at the zero radius to the boundary, based on our first-principles Grüneisen parameters under core conditions. We checked the elastic properties under different conditions from at the center of the Earth to the boundary between the inner and outer core, and find that ε-Fe is elastically isotropic under all these conditions. Thus the seismic anisotropy cannot be explained by the preferential alignments of ε-Fe in the inner core. One possible reason might be that the crystalline iron has structures other than the hcp phase in the inner core. Several different phases have been proposed before, including the double hexagonal closest packed structure, the orthorhombic phase and the bcc phase [*Andrault, et al.*, 2000; *Andrault, et al.*, 1997; *Saxena, et al.*, 1995; *Saxena, et al.*, 1993; *Yoo, et al.*, 1995]. Recent theoretical calculations show that bcc phase is extremely anisotropic to sound waves (12%) [*Belonoshko, et al.*, 2008], and becomes more stable at higher temperatures with respect to tetragonal strain [*Vocadlo, et al.*, 2008]. Most recently, it is found that fcc iron is possible stable in the inner core [*Mikhaylushkin, et al.*, 2007].

Since fcc and hcp iron have almost the same free energy under inner core conditions, it is possible tha the inner core contains both hcp and fcc iron, with changes in minor element chemistry, as well as small temperature and pressure gradients, governing the phase. This would be consistent with hetereogneities observed in the inner core seismologically [*Calvet, et al.*, 2006; *Cormier and Stroujkova*, 2005; *Ishii and Dziewonski*, 2002; *Leyton, et al.*, 2005; *Song and Helmberger*, 1998].

**III. Conclusions**

We computed first-principles elastic anisotropy for ε-Fe at high pressures and temperatures, and found that ε-Fe becomes elastically isotropic for both the compressional and shear waves under the Earth's inner core conditions. At ambient temperature, the compressional wave shows a sigmoidal shape along the propagation direction with the maximal velocity along the c axis, consistent with previous first-principles calculations. The calculated sound velocities along the Hugoniot agree with shock-compression data. Our results suggest that single phase hcp iron cannot explain the inner core anistropy after all. Several possible geophysical explanations, influence of light elements, anelasticity, and partially molten inner core, might account for the differences between the predicted elastic isotropy of $\varepsilon$-Fe under core conditions and the seismic P-wave travel time anomalies, but most likely the inner core consists of two phases, fcc and hcp iron.

**Acknowledgements**

We thank S. Y. Savrasov for kind agreement to use his LMTO codes and many helpful discussions. This work was supported by DOE ASCI/ASAP subcontract B341492 to Caltech DOE w-7405-ENG-48. Computations were performed on the Opteron Cluster at the Geophysical Laboratory and ALC cluster at Lawrence Livermore National Lab, supported by DOE and the Carnegie Institution of Washington.

1  **Figure captions**
2  Fig. 1 Compressional wave velocity $v_p$ and shear velocity $v_s$ as a function of propagation
3  direction with respect to the c-axis. Solid lines, current LMTO calculation at 50 GPa;
4  dashed lines, LAPW GGA calculation at 50 GPa [*Steinle-Neumann, et al.*, 1999; *Steinle-*
5  *Neumann, et al.*, 2001]; dotted lines, polycrystalline radial x-ray diffraction measurement
6  at 52 GPa [*Singh, et al.*, 1998].

8  Fig. 2 The agreements between the calculated compressional (solid line) and bulk
9  (dashed line) sound velocities along the shock Hugoniot, in comparison to the shock
10 compression experiment (open circles with error bars) [*Nguyen and Holmes*, 2004],
11 show evidences of the accuracy of our high-tempearture high-pressure computations.

13 Fig. 3 The temperature dependences of the elastic anisotropy for ε-Fe at 48 bohr$^3$/atom.
14 The ratio of the compressional velocity $v_p$ and two shear velocities with polarization in
15 the vertical plane $v_{s1}$ and orthogonal to the vertical plane $v_{s2}$, in respect to the maximum
16 velocities $v_{p,max}$, $v_{s1,max}$ and $v_{s2,max}$ at each given temperature are shown. The elastic
17 anisotropy decreases with increase in temperature, and essentially diminishes under
18 Earth's core temperature.

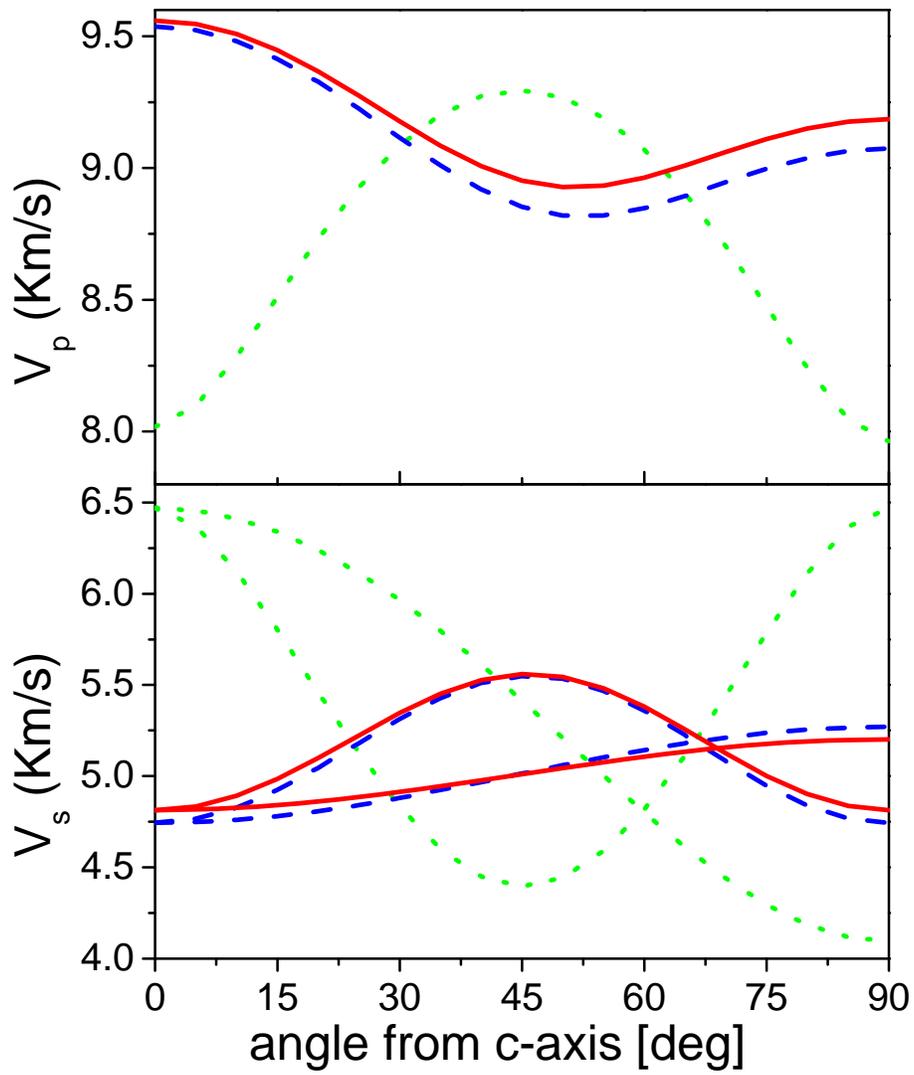

4       Figure 1

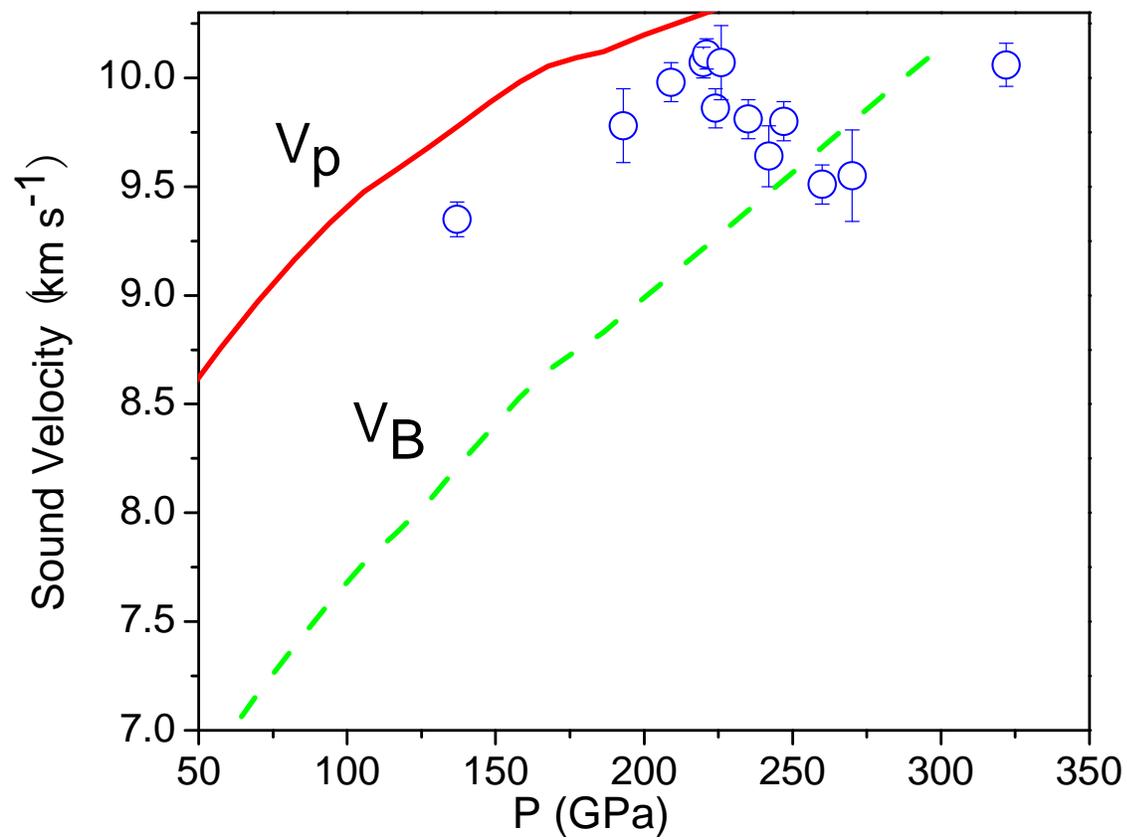

1
2
3
4 Figure 2

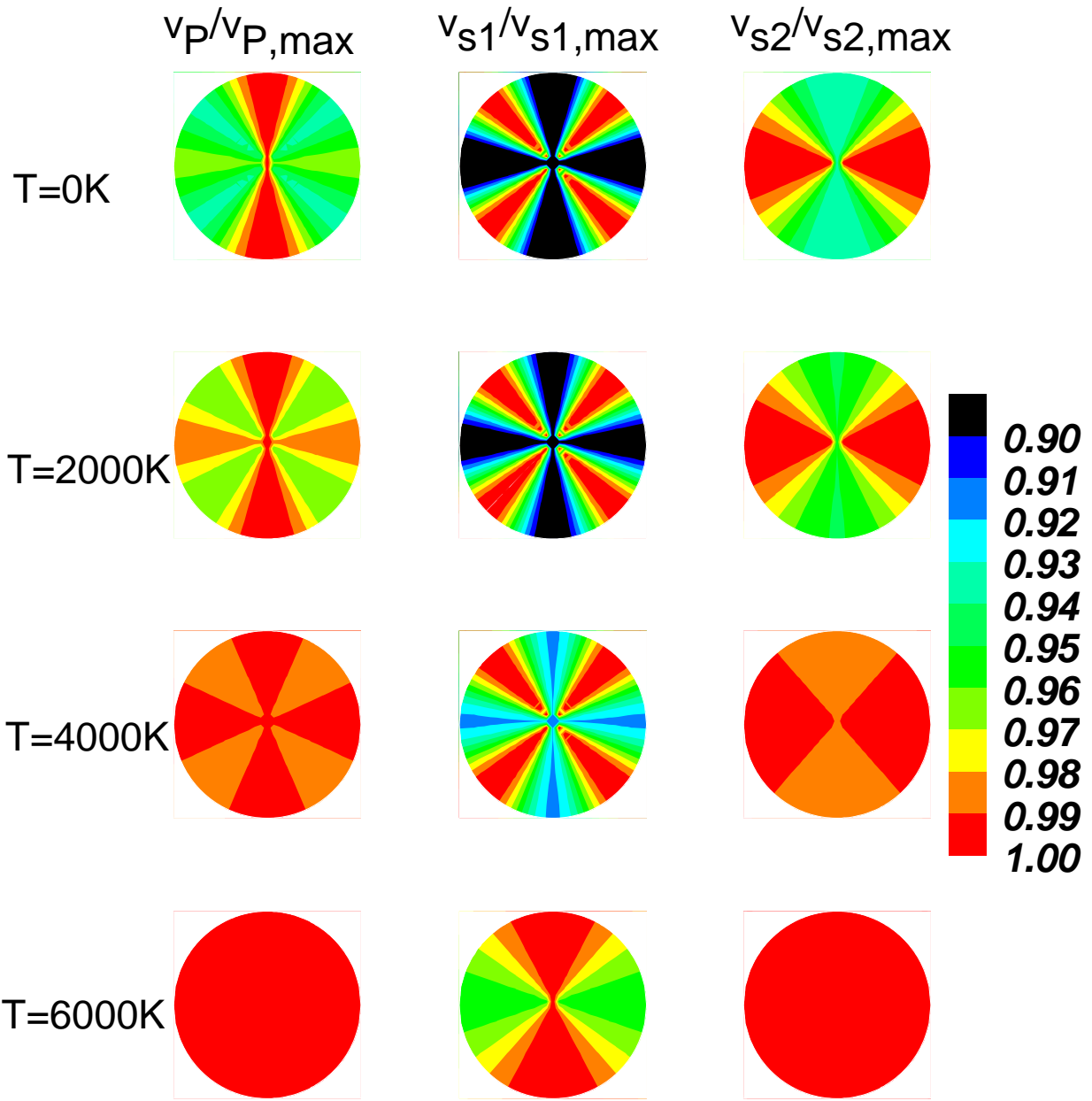

Figure 3